\documentstyle[aps,prl]{revtex}

\begin{document}
\draft
\twocolumn[\hsize\textwidth\columnwidth\hsize\csname@twocolumnfalse%
\endcsname
\title{Experimental Evidence for Resonant-Tunneling in a
Luttinger-Liquid}

\author{O. M. Auslaender$^{a}$, A. Yacoby$^{a}$, R. de Picciotto$^{b}$,
K. W. Baldwin$^{b}$, L. N. Pfeiffer$^{b}$, and K. W. West$^{b}$}
\address{
$^a$Dept. of Condensed Matter Physics, The Weizmann Institute of
Science, Rehovot, 76100, Israel\\ $^b$Bell Labs, Lucent
Technologies, 700 Mountain Ave., Murray Hill, NJ 07974, U.S.A.\\}
\maketitle
\begin{abstract}
We have measured the low temperature conductance of a
one-dimensional island embedded in a single mode quantum wire. The
quantum wire is fabricated using the cleaved edge overgrowth
technique and the tunneling is through a single state of the
island. Our results show that while the resonance line shape fits
the derivative of the Fermi function the intrinsic line width
decreases in a power law fashion as the temperature is reduced.
This behavior agrees quantitatively with Furusaki's model for
resonant tunneling in a Luttinger-liquid.
\end{abstract}
]

One-dimensional (1D) electronic systems are expected to show
unique transport behavior as a consequence of the Coulomb
interaction between carriers \cite{Fisher96}. Unlike in two and
three dimensions \cite{Altshuler85}, where the Coulomb interaction
affects the transport properties only perturbatively, in 1D it
completely modifies the ground state from its well-known
Fermi-liquid form and the Fermi surface is qualitatively altered
even for weak interactions. Today, it is well established
theoretically that the low temperature transport properties of
interacting 1D-electron systems are described in terms of a
Luttinger-liquid rather than a Fermi-liquid \cite{Kane92ab,Yue94}.
The difference between a Luttinger-liquid and Fermi-liquid becomes
dramatic already in the presence of a single impurity. According
to Landauer's theory the conductance of a single channel wire with
a barrier is given by $G=\left|t\right|^2\cdot e^2/h$, where
$\left|t\right|^2$ is the transmission probability through the
barrier. This result holds even at finite temperatures, assuming
the transmission probability is independent of energy, as is often
the case for barriers that are sufficiently above or below the
Fermi energy. In 1D, interactions play a crucial role in that they
form charge density correlations. These correlations, similar in
nature to charge density waves \cite{Glazman92}, are easily pinned
by even the smallest barrier, resulting in {\em zero} transmission
and, hence, a vanishing conductance at zero temperature. At finite
temperatures the correlation length is finite and the conductance
decreases as a power-law of temperature, $G(T)\propto T^{2/g-2}$
\cite{Kane92ab,Yue94}. Here $g\approx1/\sqrt{1+\frac{U}{2E_F}}$
where $U$ is the Coulomb energy between particles and $E_F$ is the
Fermi energy in the wire. Despite the vast theoretical
understanding of Luttinger-liquids only a handful of experiments
have been interpreted using such models. For example, in clean
semiconductor wires prepared by the cleaved edge overgrowth (CEO)
method \cite{Yacoby97}, contrary to theory, the conductance is
suppressed from its universal value \cite{Yacoby96}. Although not
fully understood this suppression is believed to be a result of
Coulomb interactions that suppress the coupling between the
reservoirs and the wire region. Other measurements done on weakly
disordered wires \cite{Tarucha95} show a weak temperature
dependence of the conductance that is attributed to the Coulomb
forces between electrons in the wire. Finally, The strongest
manifestation of interaction in the clean limit comes from
tunneling experiments such as the one recently reported on single
walled carbon nanotubes \cite{Bockrath99} and those performed on
the chiral Luttinger-liquid \cite{Chang96Grayson98}.

In this work we have focused on the transport properties through
confined states in a 1D wire, namely, when a 1D island is embedded
in a 1D wire. The 1D island is formed at low densities such that
the disorder potential in the wire exceeds the Fermi energy at
several points along the wire. Resonant tunneling (RT) has
previously been studied experimentally in a chiral
Luttinger-liquid when the resonant level width was larger than the
electron temperature \cite{Milliken96,Maasilta97}. However, an
unequivocal verification of the theoretical prediction has not
been obtained. Theoretically, the problem of RT has been
considered by Kane and Fisher \cite{Kane92ab} and was later
extended by Furusaki \cite{Furusaki98} to include many resonant
levels and the effects of Coulomb blockade (CB). Our measurements
probe the intrinsic width, $\Gamma_i$, of several resonant states
as the temperature is lowered. In contrast to conventional CB
theory \cite{Kouwenhoven96}, where $\Gamma_i$ is temperature
independent, we find that $\Gamma_i$ decreases as a power law of
temperature over our entire temperature range (2.5K to 0.25K). The
measured behavior is in quantitative agreement with the
theoretical prediction of Furusaki \cite{Furusaki98}.

The 1D wires are fabricated by the CEO method (see
Fig.~\ref{fig1}a and \cite{Yacoby97}). The electrons are confined
by a $25nm$ square well potential in one direction and by a
triangular potential well approximately $10nm$ wide (binding them
to the cleaved edge). To create a 1D island within our wire we
have chosen to study $5\mu m$ long wires that show disorder
induced deviations from the conductance plateaus (see
Fig.~\ref{fig1}b). The wire conductance is measured as a function
of its density (by negatively biasing the top gate, see
Fig.~\ref{fig1}a) using standard lock-in techniques. A fixed
excitation voltage of $10\mu V$ is applied across the wire and the
corresponding current is measured. As the density of electrons in
the wire is reduced, the 1D modes are depopulated one by one until
only a single mode remains partially populated. Several broad
resonances appear superimposed on the 'last plateau' with an
average conductance of $0.45\cdot2e^2/h$. We attribute the broad
resonances to above barrier scattering when the Fermi energy is
higher than the disorder potential. The deviation of the
conductance plateau from the universal value has been studied in
detail in \cite{Yacoby96}, however, a detailed theoretical
explanation is still lacking. It should be noted that on similar
but cleaner wires we have previously reported
\cite{Yacoby97,Yacoby96} plateau conductance values of
$\approx0.8\cdot2e^2/h$. The larger deviation observed here
suggests that disorder plays an important role in the suppression
of the value of the conductance plateaus. As the density is
further reduced, the highest potential barrier in the wire crosses
the Fermi energy, the last mode is pinched off, and the wire
splits into two parts. Upon further decrease in density a second
barrier crosses the Fermi energy, thereby, forming a 1D island,
and transport occurs through resonant states. Of course as the
density is reduced even further, more islands will form in the
wire. However, for transport to occur through them, at least one
confined state in each island must be concurrently aligned with
the Fermi energy. Since this condition is very unlikely, the wire
is expected to be completely pinched off when more than one island
has developed. Therefore, the sharp resonances in the
sub-threshold region, being almost equally spaced in gate voltage
are attributed to CB resonances through a {\em single} 1D island.

The conductance due to RT of a particle between two Fermi-liquid
leads is easily calculated using the Landauer formula,
$G_{FL}=\frac{e^2}{h}\int
\left|t(\varepsilon)\right|^2\frac{\partial
f}{\partial\varepsilon}d\varepsilon$, where
$\left|t(\varepsilon)\right|^2$ has the Breit-Wigner line shape
centered around the resonant energy $\varepsilon_0$,
$\left|t(\varepsilon)\right|^2=\frac{\Gamma_i^2}{\left(\varepsilon-
\varepsilon_0\right)^2 +\Gamma_i^2}$, and $f$ is the Fermi
function. When $k_BT\gg\Gamma_i$, the case of interest here, one
finds that $G_{FL}=\frac{e^2}{h}\Gamma_i\frac{\pi}{4k_BT}
\cosh^{-2}\left(\frac{\varepsilon_0-\mu}{2k_BT}\right)$ with $\mu$
the chemical potential in the leads. The main outcome of this
analysis is the line shape of the resonance being the derivative
of the Fermi function, its full width at half maximum equals
$3.53\cdot k_BT$, and the area under the peak (or the peak height
multiplied by $k_BT$) is proportional to $\Gamma_i$. In the
conventional theory of CB \cite{Kouwenhoven96}, $\Gamma_i$ depends
on the transmission probabilities through the individual barriers,
which are independent of temperature and hence should lead to a
peak area independent of temperature. In the case of RT from a
Luttinger-liquid, the individual transmission probabilities are
suppressed as the temperature is lowered \cite{Kane92ab,Yue94}.
Therefore it is expected and has been shown theoretically
\cite{Furusaki98} that the extracted $\Gamma_i$ should drop to
zero as $\Gamma_i\propto T^{1/g-1}$. The resonance line shape,
however, in the case of $k_BT\gg\Gamma_i$, has been shown
\cite{Maasilta97,Furusaki98} to be only slightly modified by the
interactions and the change is too small to be detected
experimentally. We, therefore, deduce the electron temperature (in
units of gate voltage) from the fit to the derivative of the Fermi
function. The deduced temperature from all the resonances in
Fig.~\ref{fig1}b is the same and follows linearly the fridge
temperature. Such a fit also allows us to calibrate the gate
voltage in units of energy thereby extracting the charging energy
(distance between peaks) estimated to be $2.2meV$. Knowing the
cross-section of the wire we estimate the length of the island
from the charging energy to be $100-200nm$. It is, therefore,
likely that the 1D island is connected on both sides to two 1D
conductors that are several microns long. Figure~\ref{fig2} shows
the extracted $\Gamma_i$ for the peaks marked in Fig.~\ref{fig1}.
It is clear that $\Gamma_i$ is not constant but rather drops as a
power law of temperature. The extracted values of $g$ for the two
peaks are 0.82 (peak $\#$1 in Fig.~\ref{fig1}) and 0.74 (peak
$\#$2 in Fig.~\ref{fig1}). The change in $g$ results from the
change in density induced in the 1D wire when moving from one peak
to the next. Similar power law behavior is observed for all
measured resonances in three different wires. The observed power
law behavior is direct proof of Luttinger-liquid behavior in our
CEO wires.

At sufficiently high temperatures the assumption of tunneling
through a single resonant state breaks down and we should expect
an increase in the extracted $\Gamma_i$ due to transport through a
few excited states of the 1D island \cite{Furusaki98}. The
possibility of an excited state affecting the temperature
dependent conductance is of interest since it allows a better test
of Furusaki's model. The excited state spectrum of the 1D island
is extracted from differential conductance measurements at finite
source drain voltage, $V_{ds}$. Figure~\ref{fig3} shows a gray
scale plot of the differential conductance as a function of the
top gate voltage and $V_{ds}$. For peak $\#$1 (in
Fig.~\ref{fig1}b) several excited states can be observed. The
lowest three, at $V_{ds}=-0.4meV$, $V_{ds}=-0.7meV$ and
$V_{ds}=-1.5meV$ are only very weakly coupled (approximately 15\%
of the intensity of the main peak) and would therefore contribute
very little to the overall conductance. However, the fourth
excited state at $V_{ds}=-1.7meV$ is more strongly coupled. Since
an excited state contributes to the conductance only when
$4k_BT\geq\Delta E$ ($\Delta E$ is the energy of the excited
state), within our temperature range of 0.25K to 2.5K only the
ground state contributes significantly and one expects a single
power law behavior as is indeed observed in Fig.~\ref{fig2}. It
should be noted that theoretically, $g$ can also be written in
terms of the charging energy, $U_c$, and the level spacing,
$\Delta E$, as $g\approx1/\sqrt{1+ U_c/\Delta E}$
\cite{Bockrath99}. In our 1D island the ratio of $U_c/\Delta
E\approx5$ and, hence, one expects $g\approx 0.4$. The large
disagreement between the measured $g$ and the expected one is not
understood at this stage.

A different case is presented in Fig.~\ref{fig4} with a strongly
coupled excited state at $V_{ds}=-0.6meV$. Hence, we expect that
at temperatures above 1.2K this excited state would contribute to
the conductance. Fig.~\ref{fig5} shows the temperature dependence
of the extracted $\Gamma_i$ of this CB peak. Indeed above 1K,
$\Gamma_i$ deviates from the low temperature power law, indicating
a contribution of an additional transport channel to the total
conductance. At low temperatures though, only the ground state
contributes to the conductance. Therefore, a fit of the low
temperature data to a power law enables us to extract a $g$ value
of 0.66 for this wire. Using this $g$ value and the measured
energy of the excited state (-0.6meV from Fig.~\ref{fig4}) we use
Furusaki's model to predict the dependence of $\Gamma_i$  over the
entire temperature range. The dashed curve in Fig.~\ref{fig5} is
the result of such a calculation where only the coupling strength
to the excited state has been adjusted. We see that the
temperature dependence predicted by the model agrees
quantitatively with the measured dependence, further supporting
the fact that Luttinger-liquid behavior describes the transport
properties of these resonances.

In conclusion we have studied the temperature dependence of CB
resonances of a 1D island embedded in an interacting 1D wire. The
observed power law behavior of the intrinsic resonance width on
temperature is direct proof of Luttinger-liquid behavior in this
system. The measured $g$ values range from 0.66 to 0.82 for the
various resonances studied. The measured behavior agrees
quantitatively with the model of Furusaki even when excited states
of the 1D island are taken into account.

We would like to thank H. L. Stormer for the fruitful
collaboration. We would also like to thank A. M. Finkel'stein and
A. Stern for helpful discussions. This work is supported by the
US-Israel BSF.

\begin{figure}
\caption{(a) - Top view layout of the wire and its contacting
scheme. The sample is fabricated using the CEO method. The 1D wire
(thick black line) exists along the cleaved surface and overlaps a
2DEG over the entire edge. The metallic gate depletes the 2DEG
over a $5\protect\mu m$ wide segment, thereby, forming an isolated
1D wire that is coupled at its ends to the overlapping 2DEG. A
further increase in the top gate voltage reduces the wire density
continuously to depletion. (b) - Conductance of the wire as a
function of the top gate voltage. Inset - A zoom-in of the
conductance of the wire in the sub-threshold region.}
 \label{fig1}
\end{figure}

\begin{figure}
\caption{The intrinsic line width of the resonance,
$\protect\Gamma_i$, vs temperature (in units of gate voltage).
Both parameters are extracted from a fit to the derivative of the
Fermi function. $\protect\Gamma_i$ is seen to decrease as a power
law of the temperature indicating Luttinger-liquid behavior. The
dashed lines are a power law fit to the data.} \label{fig2}
\end{figure}

\begin{figure}
\caption{A gray scale plot of the non-linear differential
conductance of peak \protect$\#$1 (see Fig.~\protect\ref{fig1}b).
$V_{ds}$ is stepped with $100\mu V$ intervals.} \label{fig3}
\end{figure}

\begin{figure}
\caption{A gray scale plot of the non-linear differential
conductance of a resonance that has a strongly coupled state at
$V_{ds}=-0.6meV$. $V_{ds}$  is stepped with $20\mu V$ intervals.}
\label{fig4}
\end{figure}

\begin{figure}
\caption{The intrinsic line width of the resonance described in
Fig.~\protect\ref{fig4} vs temperature (in units of gate voltage).
The dashed line is a fit to the data based on Furusaki's model.
$g$ is determined from the low temperature behavior and the energy
of the excited state is determined from Fig.~\protect\ref{fig4}.
The coupling strength to the excited state is the only adjustable
parameter in the fit.} \label{fig5}
\end{figure}

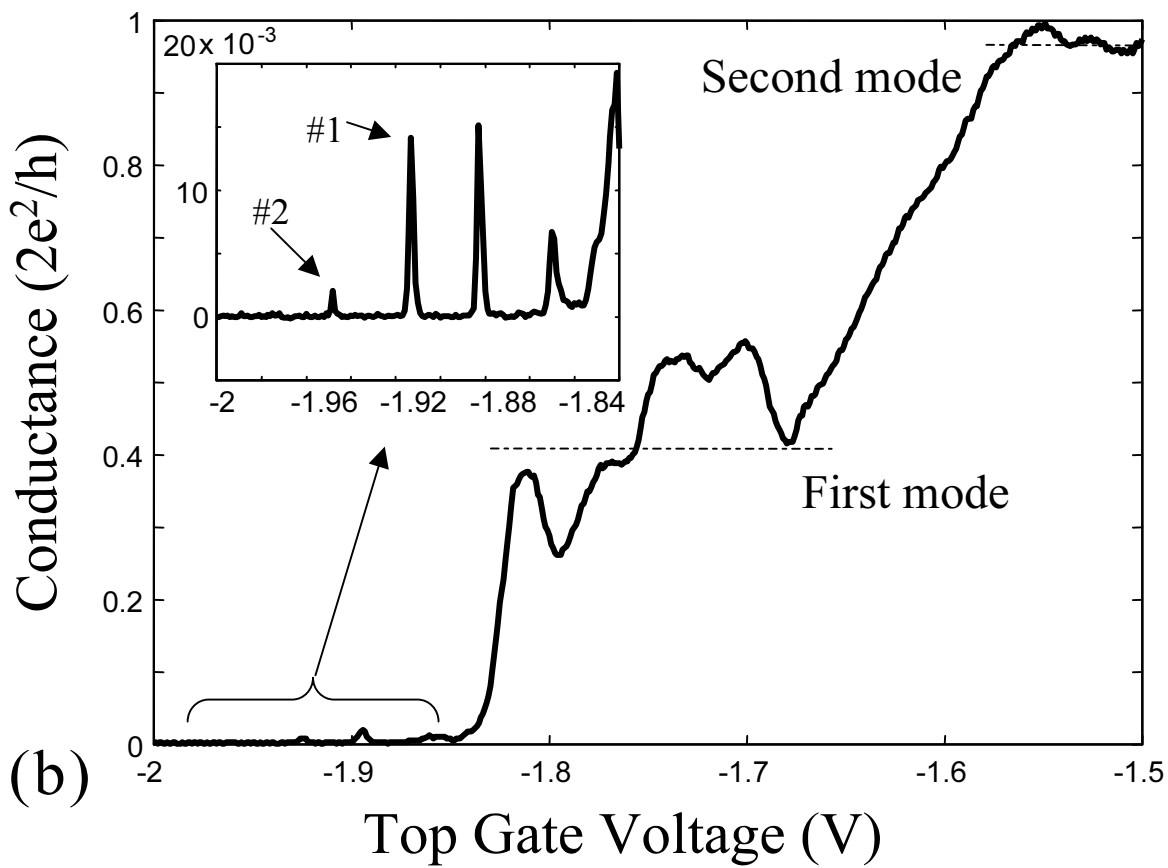
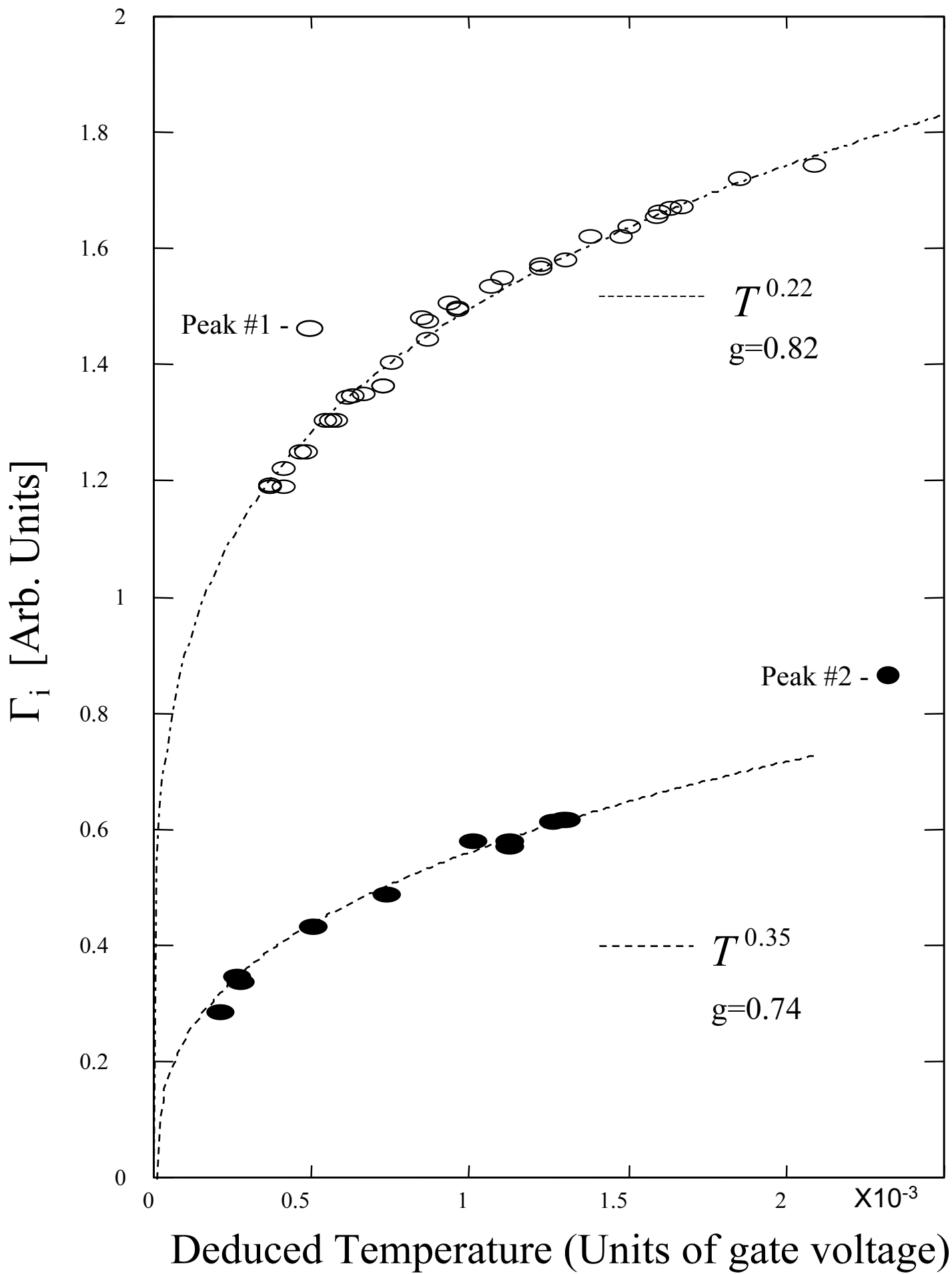
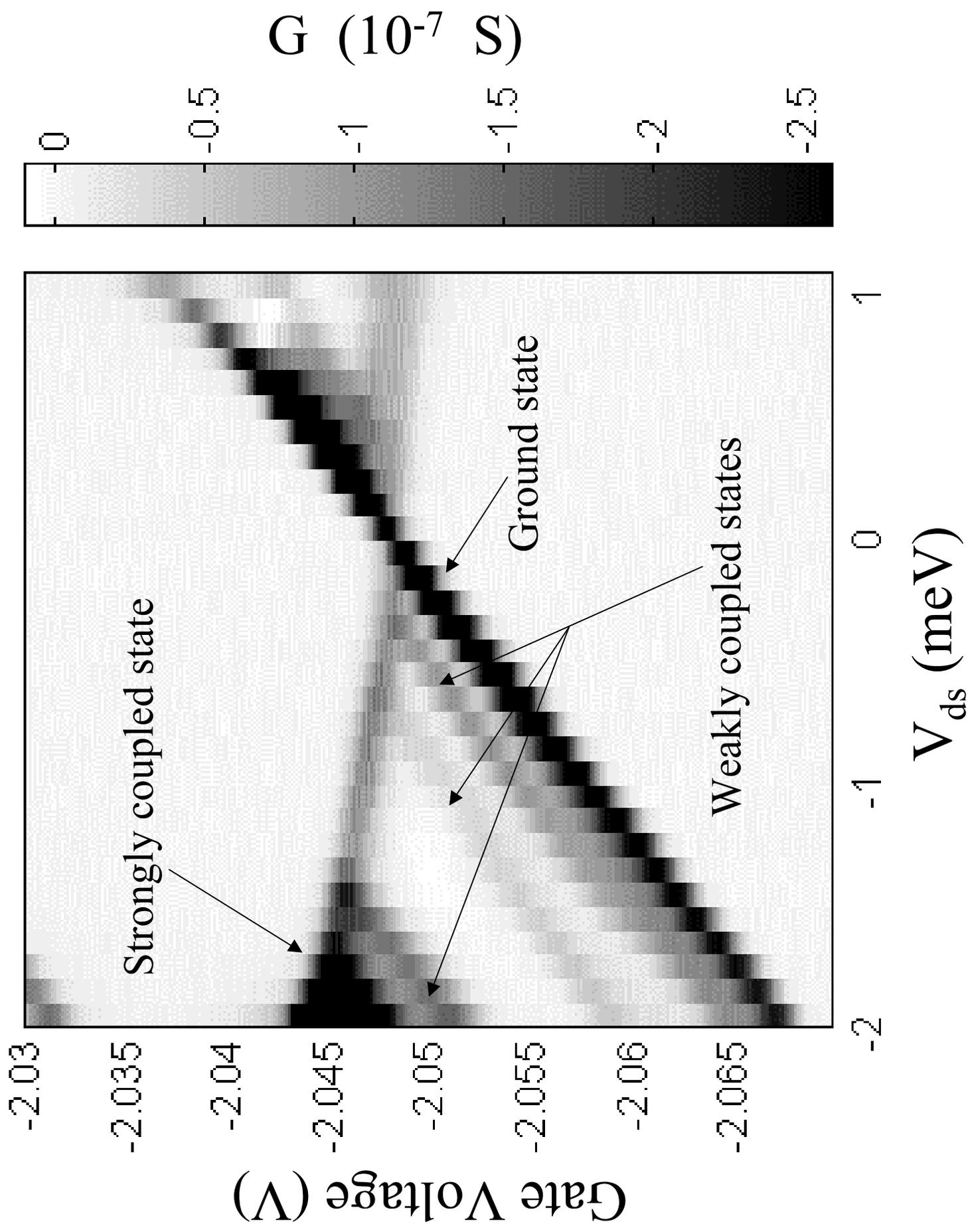
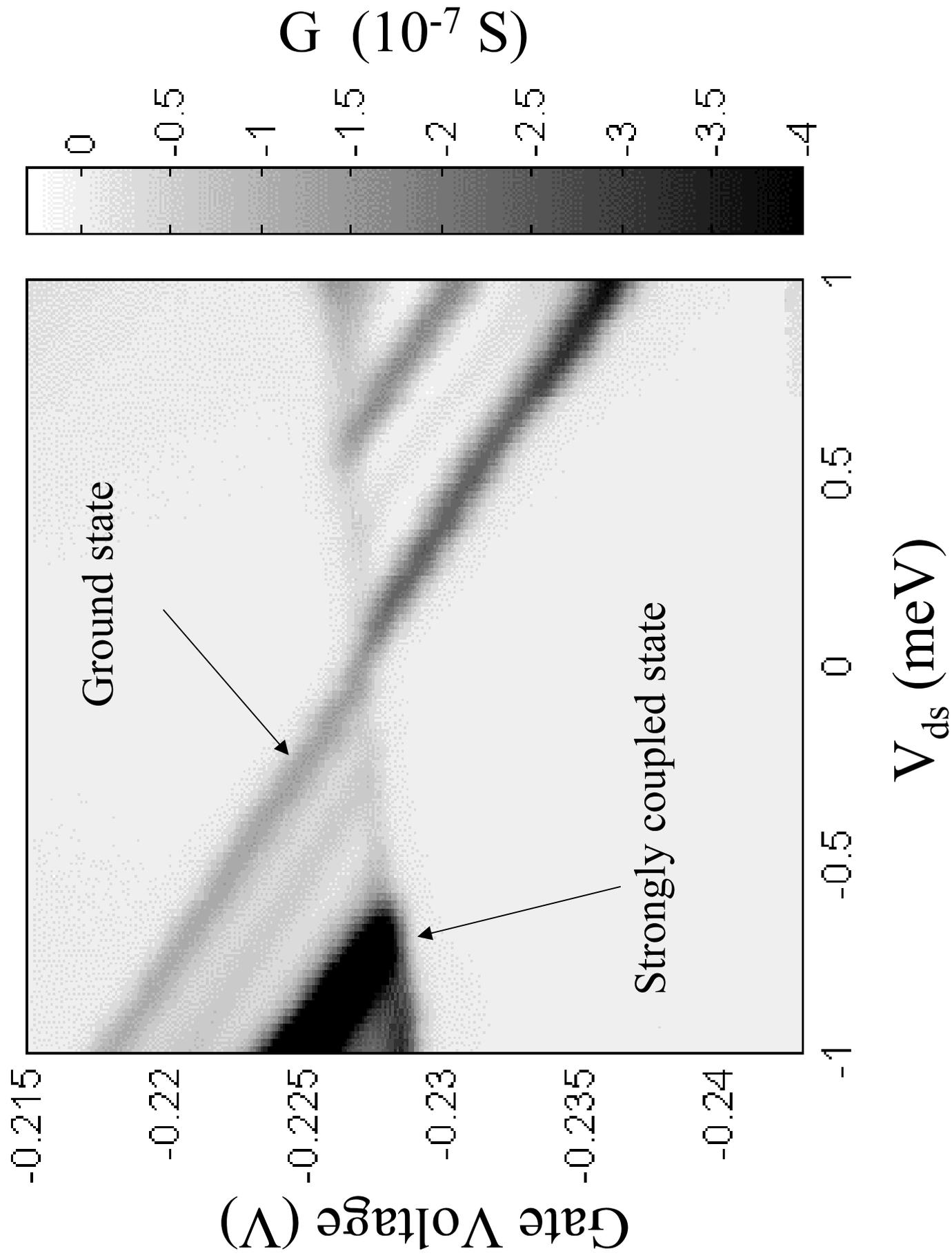
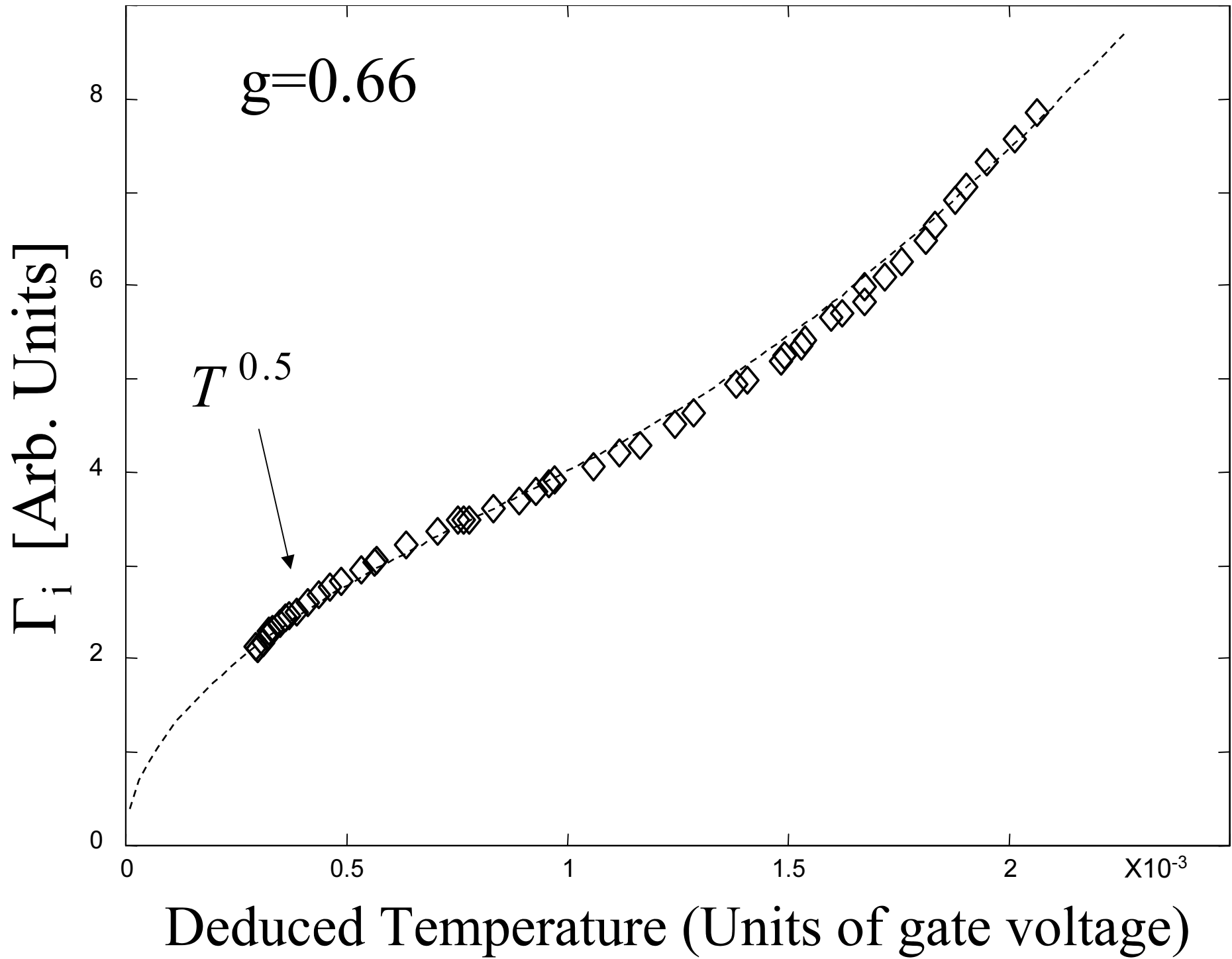
\end{document}